\documentclass[12pt,preprint]{aastex}

\begin{document}
\pagenumbering{arabic}

\title{WHAT ARE S0 GALAXIES?}

\author{Sidney van den Bergh}
\affil{Dominion Astrophysical Observatory, Herzberg Institute of Astrophysics, National Research Council of Canada, 5071 West Saanich Road, Victoria, BC, V9E 2E7, Canada}
\email{sidney.vandenbergh@nrc-cnrc.gc.ca}

\begin{abstract}  

 The data collected in the Shapley-Ames catalog of bright
galaxies show that lenticular (S0) galaxies are typically about a
magnitude fainter than both elliptical (E) and early spiral
(Sa) galaxies. Hubble (1936) was therefore wrong to regard S0
galaxies as being intermediate between morphological types E 
and Sa. The observation that E5-E7 galaxies are significantly fainter
than objects of sub-types E0-E5 suggests that many of the flattest
``ellipticals" may actually be misclassified lenticular galaxies. In
particular it is tentatively suggested all E7 galaxies might actually 
be misclassified S0$_{1}$(7) galaxies. The present results are consistent 
with the view that galaxies belonging to the S0 class evolved in 
environments in which they typically lost more than half of their 
original luminous material. 

\end{abstract}


\section{INTRODUCTION}

In his pioneering paper on the classification of galaxies
(Hubble 1926) wrote ``The complete series [of ellipticals]
is E0, E1,....,E7, the last representing a definite limiting 
figure which marks the junction with spirals.'' Later Hubble
(1936, p.45) introduced, what he referred to as a more or less
hypothetical stage S0 between the elliptical and spiral galaxies.
Hubble (1936, p.42) also noted that the transition from ``types E7 
to Sa may be cataclysmic - all known examples of Sa have
fully developed arms.'' The definition of the S0 class was later 
updated and refined by Sandage (1975), by Sandage \& Tammann 
(1981)  and by Sandage \& Bedke (1994). It is the purpose of the
present investigation to make two points: (1) Both E and Sa 
galaxies are typically more luminous than S0 galaxies, so that 
lenticulars cannot be regarded as being truly intermediate between
elliptical and spiral galaxies. (2) Galaxies of sub-types E5-E7
appear to be significantly fainter than those of sub-types E0-E4.
This anomaly might be accounted for  by assuming that a
significant fraction of the flattest ``ellipticals'' are actually 
misclassified  lenticulars.  In particular  it is tentatively suggested 
that all E7 galaxies might actually be such  misclassified S0 galaxies. 
If this conjecture is correct then one needs to reexamine Hubble's 
(1926) claim that: `` The complete series is E0, E1,....,E7, the last 
representing a definite limiting figure which marks the junction 
with the spirals." The data that now exist
appear to suggest that S0 galaxies are actually early-type galaxies 
that, because of the environment in which they were located, lost 
a significant fraction of their original luminous material. 

\section{DATABASE}

 The data on E,  S0  and Sa galaxies used in the present study were 
drawn from the 1276 galaxies listed in {\it A Revised Shapley-Ames Catalog of Bright Galaxies} (Sandage \& Tammann 1981), which is almost entirely based on the classification of galaxy images on plates obtained with large reflecting telescopes that were classified, in a uniform manner, by expert morphologists. The listed M$_{B}$ magnitudes are the values

        $M^{o,i}_{B_{T}}$
  
{\flushleft given by Sandage \& Tammann. Note that $M_{B}$ is corrected for both internal and foreground absorption.  These $M_{B}$ values were based on the assumption that H$_{o}$ = 50 km s$^{-1}$ Mpc$^{-1}$. Since it is now believed that the Hubble parameter is actually $\sim$70 km s$^{-1}$ Mpc$^{-1}$ (Komatsu et al. 2009) the luminosity of the Local Group galaxy NGC 221 (M32) 
was, for the sake of consistency, increased by 0.73 mag to bring it 
onto the Sandage \& Tammann luminosity scale. To simplify the analysis  objects of intermediate morphological types, such as E/S0, S0, S0/Sa have been excluded from the present study. Also dropped from the present sample were two galaxies that Sandage \& Tammann classified as dE. It is often difficult (or impossible) to interpret the flattening values 10(a-b)/a of barred galaxies. It was therefore decided to also excluded objects of types SB0 and SBa from the present study of the frequency and distribution galaxy flattening values.}

\section{ DISCUSSION}

 The luminosity distribution of the E, S0 and Sa galaxies in the sample defined in Section 2, is  presented in Table 1 and is plotted in Figure 1. The most unexpected feature revealed by these data  is 
that S0 galaxies are, on average, significantly fainter than both elliptical galaxies and spirals of morphological type Sa. In other words S0 galaxies are {\it not} intermediate between types E and Sa. A Kolmogorov-Smirnov test shows that the probability that the E and S0 galaxies in the present sample were drawn from the same parent luminosity distribution is $<$0.01\%. By the same token, the probability that the S0 and Sa samples were drawn from the same parent populations is found to be $\sim$0.1\%. On the other hand an S-K test  can not reject the hypothesis that the E and Sa galaxies in the Shapley-Ames catalog were  drawn from the same parent distribution of luminosities. The data in Table 1 may be interpreted in two different ways: Either (1) the  lenticular galaxies  contain a  population component, peaking at M$_{B}$ $\simeq$ --19.7 mag,  that is not present among E and Sa galaxies or, perhaps more plausibly, (2) S0 galaxies lived (or were assembled in) an environment in which typically more than half of their initial luminous material was lost. Plausible mechanisms to account for such mass removal were first proposed by Spitzer \& Baade (1951) and by Gunn \& Gott (1972). Anemic spirals, such as NGC 4921 in the Coma cluster (van den Bergh 1991) may be examples of objects that are presently being stripped of gas and that are expected to eventually turn into lenticular galaxies. From the data in Table 1 it is seen that the median luminosity of E galaxies is M$_{B}$ = --21.4 mag, compared to a median luminosity of  M$_{B}$ = --20.4 mag for S0 galaxies.  Could this difference have been caused by S0 galaxies having dropped out of the sample more easily than E's because of internal absorption?  The answer to this question is provided by Sandage \& Tammann (1978), who studied the inclination dependence of the colors of lenticular galaxies.  From this investigation those authors found that the optical depth in V for dust extinction is less than $\tau$=0.01 for an average S0 galaxy.   The median luminosity of the Sa galaxies in the present sample is  M$_{B}$ = --21.2 mag, which is also significantly brighter than the corresponding value for lenticular galaxies. These results suggest that S0 galaxies typically lost slightly over half of their initial luminous material during the course of their evolution. 
   
 Even though the numbers involved are not large, the data in Table 2 suggest the possibility that the luminosity distribution of elliptical galaxies in the Shapley-Ames catalog is a function of their apparent flattening. The most flattened ellipticals with morphological sub-types E5-E7 are, on average, seen to be less luminous than are the rounder ellipticals of sub-types E0-E4. This effect is seen even more clearly in the 2x2 contingency table that is shown as Table 3. A Chi-squared test of the counts in this table shows that there is $<$0.1\% probability that the E0-E4 and E5-E7 galaxies were drawn from the same parent population. It is particularly noteworthy that the two only E7 galaxies in the Shapley-Ames catalog (NGC 4342 with M$_{B}$ = --18.16 mag and NGC 4623 with M$_{B}$ = --18.61 mag are both exceptionally faint. Since lenticular  galaxies are significantly less luminous that ellipticals  it is tempting to speculate that this effect is due to the fact that many of the flattest elliptical galaxies having Hubble sub-types E5-E7 are actually misclassified lenticulars of types S0(5)-S0(7). This suspicion is strengthend by the fact that NGC 3115, which Hubble(1936, p. 91) used as the proto-type for subclass E7, has more 
recently been assigned to type S0$_{1}$(7) by Sandage \& Tammann (1981). If this conclusion is correct, then true ellipticals  only span the range E0-E6, rather than the range E0-E7 originally assigned to these objects by Hubble. The following E5 galaxies in the Shapley-Ames
catalog are so faint that they might also turn out to be be misclassified S0 galaxies: NGC 3156 (M$_{B}$ = --18.50 mag),  NGC 3605 (M$_{B}$= --17.29 mag) and NGC 4476 (M$_{B}$ = --18.62 mag).
  
 Among lenticular galaxies of sub-types S0(1) to S0(8) luminosity does not appear to depend on flattening. However, the 11 very flat galaxies that Sandage \& Tammann assign to subtypes  S0(9) and S0(10) appear to be significant fainter than the less flattened objects of of subclasses S0(0) to S0(8). Since almost all of these objects are of the dustless sub-type S0$_{1}$, it appears unlikely that the faintness of the very flattest lenticulars is due to dust absorption. The work of Sandage \& Visvanathan (1978) shows that the optical depth for an average S0 galaxy is less than $\tau$=0.01.   A K-S test  shows that there is only a $\sim$1\% probability that these two classes of objects were drawn from the same parent population of luminosities. This result may indicate that super-flat lenticulars are particularly prone to the loss of luminous material during the course of their evolution.

It is noted in passing that (1) the luminosity distribution of Shapley-Ames S0 galaxies does not differ systematically from that of those of type SB0, and (2) that there appear to be no systematic differences between the luminosity distributions of objects of types S0$_{1}$, S0$_{2}$ and S0$_{3}$.

 One might ask if the observed difference in the luminosity 
distributions of E, S0 and Sa galaxies might be a consequence of 
some distance-dependent selection effect. To look into this question 
Table 1 lists  results separately for  131 relatively nearby Shapley-
Ames galaxies with $(m-M)_{o}$ $<$ 32.5 and for 199 more distant 
bjects having $(m-M)_{o}$ $\geqslant$32.5. These data show that E + Sa 
galaxies are more luminous than S0 galaxies in both of these 
distance ranges.  A K-S test shows  that this result has a confidence  
level of  91\% for the nearby galaxies and and of 99.1\% for the 
more distant Shapley-Ames sub-sample.

   Finally, Table 4 shows the frequency distribution of morphological 
types among Shapley-Ames galaxies as a function of their distance. 
These data show that S0 galaxies have a much higher frequency 
among nearby galaxies than they do among distant ones. 
Furthermore the numbers in Table 4 show that this effect persists, 
even when the members of the Virgo cluster, which contains many 
lenticular galaxies, is excluded from the data. These results can
be interpreted in two ways. Either (1) a significant number of 
distant objects of type S0 have been misclassified as ellipticals or, 
perhaps more plausibly, (2) S0 galaxies are underrepresented in the 
distant sample because of their low luminosity.  In other words, the result may have been affected by Malmquist bias.  Due to rounding errors the columns in Table 4 do not all add up to 100\%.

\section{CONCLUSIONS}

It is suggested that  S0 galaxies are not, as Hubble originally suggested, transitional objects between classes E and S0. Rather,
their low luminosities  may indicate that they are actually galaxies which lived in environments that typically caused them to lose more than half of their initial luminous material. The low luminosity
of the small number of galaxies  classified as E7 indicates that these objects  might actually be misclassified lenticulars of  type S0(7). 
If so, true elliptical galaxies only span the morphological range E0 to E6. It is also found that the flattest lenticulars of sub-types S0(9) and S0(10) are significantly fainter than the less flattened objects 
of sub-types S0(0) to S0(8).

 My interest in the subject matter of this paper was sparked by a colloquium given at the Princeton University Observatory by Walter Baade in 1949 or early 1950. Towards the end of this lecture Lyman Spitzer raised this arm and suggested that the S0 galaxies in dense clusters might have been formed as a result of removal of gas during collisions between spiral galaxies in dense clusters.

Thanks are due to Stefano Andreon who urged me to look into possible distance-dependent effects on the morphological classifications of E, S0 and Sa galaxies.  I am also indebted to the referee for pointing out that reference should be made to the possible effects of internal absorption on the sample of S0 galaxies.  I am also indebted to Brenda Parrish and Jason Shrivell for technical assistance.

%\clearpage

\begin{deluxetable}{rrrrrrrrrr}
\tablewidth{0pt}   
\tablecaption{Luminosity distribution of early-type Shapley-Ames galaxies}
\tablehead {& \multicolumn{3}{c}{All distances} 
& \multicolumn{3}{c}{$(m-M)_{o} < 32.5$} &  \multicolumn{3}{c}{$(m-M)_{o}$ $\geqslant 32.5$}\\
 \colhead{M$_{B}$} & \colhead{N(E)} & \colhead{N(S0)} 
& \colhead{N(Sa)} & \colhead{N(E)} & \colhead{N(S0)} & \colhead{N(Sa)}
& \colhead{N(E)} & \colhead{N(S0)} & \colhead{N(Sa)}}

\startdata

-23.00 to -23.49 &  1  & 0 & 1  & 0   & 0 &  1  & 1  & 0   & 0 \\
-22.50~~~~    -22.99 &  12 & 1 & 2  & 0   & 0 &  0  & 12 & 1   & 2 \\
-22.00~~~~    -22.49 &  12 & 4 & 10 & 1   & 0 &  0  & 11 & 4   & 10 \\
-21.50~~~~    -21.99 &  37 & 11 & 10 & 2  &  3 &  0 & 35 &  8  & 10 \\
-21.00~~~~    -21.49 &  31 & 16 &  22 & 5 &  2 & 10 & 26 & 14  & 12 \\
-20.50~~~~    -20.99 &  20 & 17 & 12  & 6 &  7 &  6 &  14 & 10 & 6 \\
-20.00~~~~    -20.49 & 10  & 16 &  12 & 6 & 12 &  9  &  4 &  4 & 3 \\
-19.50~~~~    -19.99 & 12  & 24 &  6  & 9 & 17 &  6  &  3 &  7 & 0  \\
-19.00~~~~    -19.49 & 4   & 11 &  2  & 4 & 10 &  2  &  0 &  1 & 0 \\
-18.50~~~~    -18.99 & 4   & 2  &  0  & 3 & 2  &  0  &  1 &  0 & 0 \\
-18.00~~~~    -18.49 & 1   & 3  &  0  & 1 & 3  &  0  &  0 &  0 & 0  \\
-17.50~~~~    -17.99 & 0   & 0  &  0  & 0 & 0  &  0  &  0 &  0 & 0 \\
-17.00~~~~    -17.49 & 1   & 1  &  0  & 1 & 1  &  0  &  0 &  0 & 0 \\
-16.50~~~~    -16.99 & 0   & 0  &  0  & 0 & 0  &  0  &  0 &  0 & 0 \\
-16.00~~~~    -16.49 & 1   & 1  &  0  & 1 & 1  &  0  &  0 &  0 &  0 \\

Total &   146    &  107 & 77 &  39 &  58 & 34 & 107 &  49  & 43\\
\enddata
\end{deluxetable}

%\clearpage

\begin{deluxetable}{rcc}
\tablewidth{0pt}   
\tablecaption{Luminosity distribution among elliptical galaxies}
\tablehead{\colhead{M$_{B}$} & \colhead{N(E0-E4)} & \colhead{N(E5-E7)}} 

\startdata

-23.00 to  -23.49  &      1        &        0\\
-22.50~~~~     -22.99  &     10        &        2\\
-22.00~~~~     -22.49  &     11        &        1\\
-21.50~~~~     -21.99  &     31        &        6\\
-21.00~~~~     -21.49  &     25        &        6\\
-20.50~~~~     -20.99  &     17        &        3\\
-20.00~~~~     -20.49  &      8        &        2\\
-19.50~~~~     -19.99  &     10        &        2\\
-19.00~~~~     -19.49  &      2        &        2\\
-18.50~~~~     -18.99  &      1        &        3\\
-18.00~~~~     -18.49  &      0        &        1\\
-17.50~~~~     -17.99  &      0        &        0\\
-17.00~~~~     -17.49  &      0        &        1\\
-16.50~~~~     -16.99  &      0        &        0\\
-16.00~~~~     -16.49  &      1        &        0\\

\enddata
\end{deluxetable}

\begin{deluxetable}{crr}
\tablewidth{0pt}   
\tablecaption{Flattening-luminosity contingency table}
\tablehead {& \colhead{E0-E4} & \colhead{E5-E7}}

\startdata

M$_{B}$ $\lesssim$ -19.0  &  115   &   24\\
M$_{B}$ $>$  -19.0    &    2   &    5\\

\enddata
\end{deluxetable}

\begin{deluxetable}{rrrr}
\tablewidth{0pt}   
\tablecaption{Comparison between the frequency distributions of Hubble types among nearby {$[(m-M)_{o} < 32.5$}]and distant {[$(m-M)_{o}$ $\geqslant 32.5$]}}
\tablehead {\colhead{Type} & \colhead{Nearby} & \colhead{Nearby} & \colhead{Distant}\\
& &  \colhead{(Virgo excluded)}}

\startdata

E  & 30\% & 28\% & 54\% \\
S0 & 44   & 42   & 25 \\
Sa & 26   & 30   & 22 \\

\enddata
\end{deluxetable}

\clearpage

\begin{figure}
\epsscale{.6}
\plotone{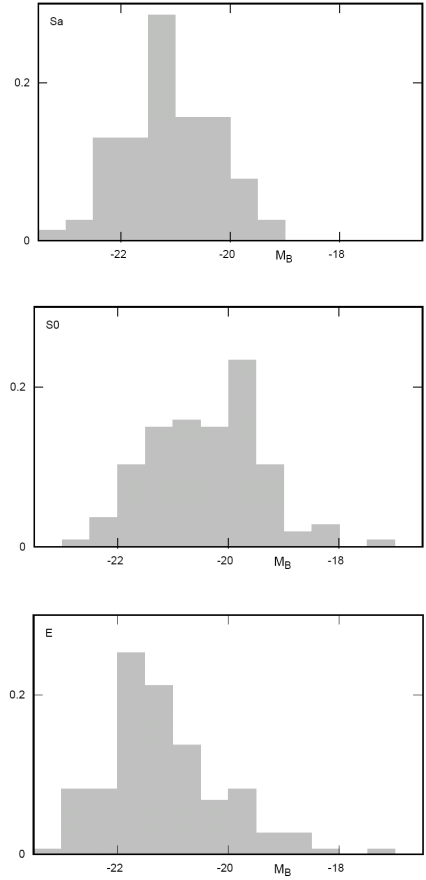}
\caption{Normalized luminosity distributions for E,S0 
and Sa galaxies in the Shapley-Ames catalog. The figure shows that S0 
galaxies are, on average, significantly fainter than E and Sa 
galaxies.  Lenticulars are therefore not strictly intermediate between 
ellipticals early-type spirals}
\end{figure}

\end{document}